# libterachem.py: Modular development of electronic structure methods with graphical processing units

Juncheng Harry Zhang[1,2], Ruiyan Wang[1,2], Andreas Erbs Hillers-Bendtsen[1,2], Todd J. Martínez[1,2,*]

[1]Department of Chemistry and the PULSE Institute, Stanford University, Stanford, California 94305, USA

[2]SLAC National Accelerator Laboratory, Menlo Park, California 94025, USA

* Corresponding author: todd.martinez@stanford.edu.

**Abstract:** We present `libterachem.py`, a Python interface and method framework that facilitates development and prototyping of electronic structure methods while leveraging TeraChem's optimized graphical processing unit (GPU) operations. Direct access to these building blocks and their intermediate quantities allows users to assemble and modify methods within Python while reusing the underlying GPU implementations. A layered architecture, programmable at both the operation level and the method level, supports numerical experimentation, composition of complete electronic structure methods with interchangeable components, and integration with external scientific libraries. We illustrate operation-level use with grid pruning for least-squares tensor hypercontraction and with exchange–correlation functionals supplied through LibXC and by the machine-learned Skala functional. We illustrate method-level use with dispersion corrections, periodic calculations, and geometry optimization with an external optimizer. Python-driven self-consistent-field calculations using `libterachem.py` incur no significant computational overhead compared to native TeraChem and preserve its multi-GPU scaling. A photoactive yellow protein QM/MM molecular dynamics application further demonstrates throughput comparable to native TeraChem in a workflow composed with external simulation libraries. These results establish `libterachem.py` as a platform for method development and interoperable molecular simulation that retains the performance of TeraChem's GPU operations.

## I. Introduction

Building an electronic-structure method often requires modifying selected numerical steps and inspecting the quantities they produce, while reusing established, optimized operations. Many of these operations are shared across methods. Coulomb (J) and exchange (K) matrix builds, for example, underlie Hartree–Fock and Kohn–Sham theory and modern formulations of both single-reference[1-4] and multiconfigurational[5-8] excited-state methods. Ideally, one could change how a method combines these operations or introduce a new approximation without rewriting every building block. Graphical processing units (GPUs) make this separation more valuable, because optimizing an operation for the GPU is specialized work whose implementation can serve many methods.[9] The same goal of modularity goes beyond the electronic structure calculation. One desires simple composition of a method with independently developed corrections, molecular environments, and simulation drivers. Python is a natural setting in which to combine such operations with numerical analysis and with established scientific libraries.

Assembling methods in Python leveraging compiled operations is by now established practice. PySCF exposes the individual steps of a calculation so that a method can be assembled and modified in Python.[10] Psi4NumPy builds reference implementations on Psi4's compiled operations,[11] GPU4PySCF carries this style onto the GPU,[12] and VeloxChem shows the same layered Python and C++ organization.[13] The operations need not live inside one package. GauXC demonstrates that a specialized GPU operation, Fock construction, can be shared as a library across programs.[14] Machine learning now enters at several levels, from learned interatomic potentials that replace the electronic calculation at inference,[15] to neural-network wavefunctions optimized within an electronic structure calculation,[16] to learned exchange–correlation (XC) functionals that replace one component of a Kohn–Sham calculation.[17-18] These models are built in machine-learning frameworks, so combining one with optimized grid and integral operations is simplest when calling them from Python. Two capabilities recur across these efforts: driving the steps inside one calculation, and composing complete calculations into larger workflows. In both cases, what a developer can build is bounded by the operations reachable from Python.

TeraChem was the first electronic-structure package designed from the ground up to exploit GPUs,[19-22] with its optimized operations encapsulated in self-contained libraries called boxes.[23] This organization has already facilitated method development inside the compiled code. For example, GPU-accelerated J and K construction is provided using either ab initio integrals (IntBox) or semiempirical integrals (SQMBox). The use of a common interface for SQMBox and IntBox automatically enables semiempirical extensions of multiconfigurational methods[24] whose ab initio variants have already been implemented. Up until now, there were two ways to access TeraChem operations from Python. The TeraChem Protocol Buffers (TCPB) interface packages a complete calculation (e.g. energy or energy-and-gradient) and submits it to a running TeraChem server, possibly on another computer.[25] TeraChem Cloud, and its successors BigChem and ChemCloud, distribute many such calculations across GPU workers and compose them into workflows.[26-27] These server and cloud interfaces operate on complete calculations. Lightspeed took the opposite approach, exposing box-level operations from TeraChem's GPU integral and configuration-interaction (CI) boxes to Python and assembling methods there.[28] This approach supported prototyping and development of new electronic structure methods.[29-31] It was interfaced to the integral and CI components of TeraChem (intbox and cibox). Up until now, no existing interface provided operation-level access to TeraChem's broader set of boxes as an installable package.

In this work we present `libterachem.py`, which exposes TeraChem's production GPU operations and their intermediate quantities directly within Python and provides a framework for assembling them into complete electronic structure methods. NVIDIA's recent cuEST is a closely related effort, exposing GPU electronic-structure operations for use in caller-defined calculations.[32-33] Such efforts reflect a growing interest in reusable GPU building blocks, a long-standing organizing principle of TeraChem's boxes.[23] `libterachem.py` enables easy incorporation of TeraChem's computational capabilities into Python code, including multiple J and K engines,[19, 34] least-squares tensor hypercontraction (THC) fitting,[35-36] configuration interaction,[37-38] and Γ-point periodic integrals.[39] These operations support three uses that organize the paper: direct experimentation with numerical constructions and their intermediate quantities, assembly of complete methods in a reusable Python framework, and composition with external libraries.

These uses rest on two distinctions in the architecture (Figure 1, Section II). The first is the level at which a developer works. At the *operation level*, a developer calls the operations directly as building blocks and combines them with Python array computations to write or alter a method. At the *method level*, the `tc` framework assembles the same operations into a complete method that is selected and run as a whole. The second distinction separates what TeraChem implements from what external libraries contribute. A small integration layer, `tcx`, adapts external libraries at either level. An external library can thus supply an operation (e.g., the evaluation of an exchange-correlation functional) used by the *method level* or it can call a routine at the *method level* (e.g.

geometry optimization or molecular dynamics which would obtain energies and forces from a method such as DFT). Researchers can work at the level a change requires while reusing the implementation below it.

The examples in Section III demonstrate this control, from the operation level to complete methods composed with external libraries. We investigate grid pruning and fitting-matrix conditioning in least-squares THC, and supply XC energies and derivatives through PyLibXC and through the machine-learned Skala functional, whose potential is obtained by automatic differentiation, while retaining TeraChem's grid and integral operations.[18, 40] Method-level examples combine complete methods with dispersion corrections, periodic calculations, and external geometry optimization.

We assess the computational cost of this flexibility in Section IV. Python-driven SCF iterations achieve comparable performance to native TeraChem and preserve its multi-GPU scaling. A complete photoactive yellow protein QM/MM molecular dynamics workflow, combining TeraChem, OpenMM,[41] and ASE,[42] likewise achieves throughput comparable to native implementations in TeraChem.

## II. Software architecture

An electronic structure method computes the energy and related properties of a molecular system from its geometry and composition. Methods differ widely in the physics they capture, from mean-field theories to correlated wavefunctions, yet are built from a small set of recurring operations. Nearly all evaluate molecular integrals. Most form a mean-field potential and drive it to self-consistency. Many add numerical quadrature and XC evaluation, a configuration-interaction step, or the matching nuclear gradients. TeraChem exposes many of these recurring operations as reusable building blocks, so a method is assembled by combining operations rather than reimplementing each one.

`libterachem.py` organizes access to these building blocks at two levels, the individual operations and the complete methods built from them, through a small set of Python modules (Figure 1). The self-contained libraries that carry out the individual electronic structure operations (TeraChem's "boxes") form the lowest level. Two modules present these operations to Python: a machine-generated interface, `terachem_box`, makes each box operation callable, and a handwritten utility, `terachem_util`, supplies the setup, bookkeeping, and shared abstractions that a method requires. Together they form the low-level interface, where the operations can be used directly and combined with array libraries such as NumPy or CuPy. Building on these two modules, the *method level* framework `tc` assembles the operations into complete electronic structure methods. The integration layer `tcx` spans both levels, adapting external libraries to the interfaces the other modules define.

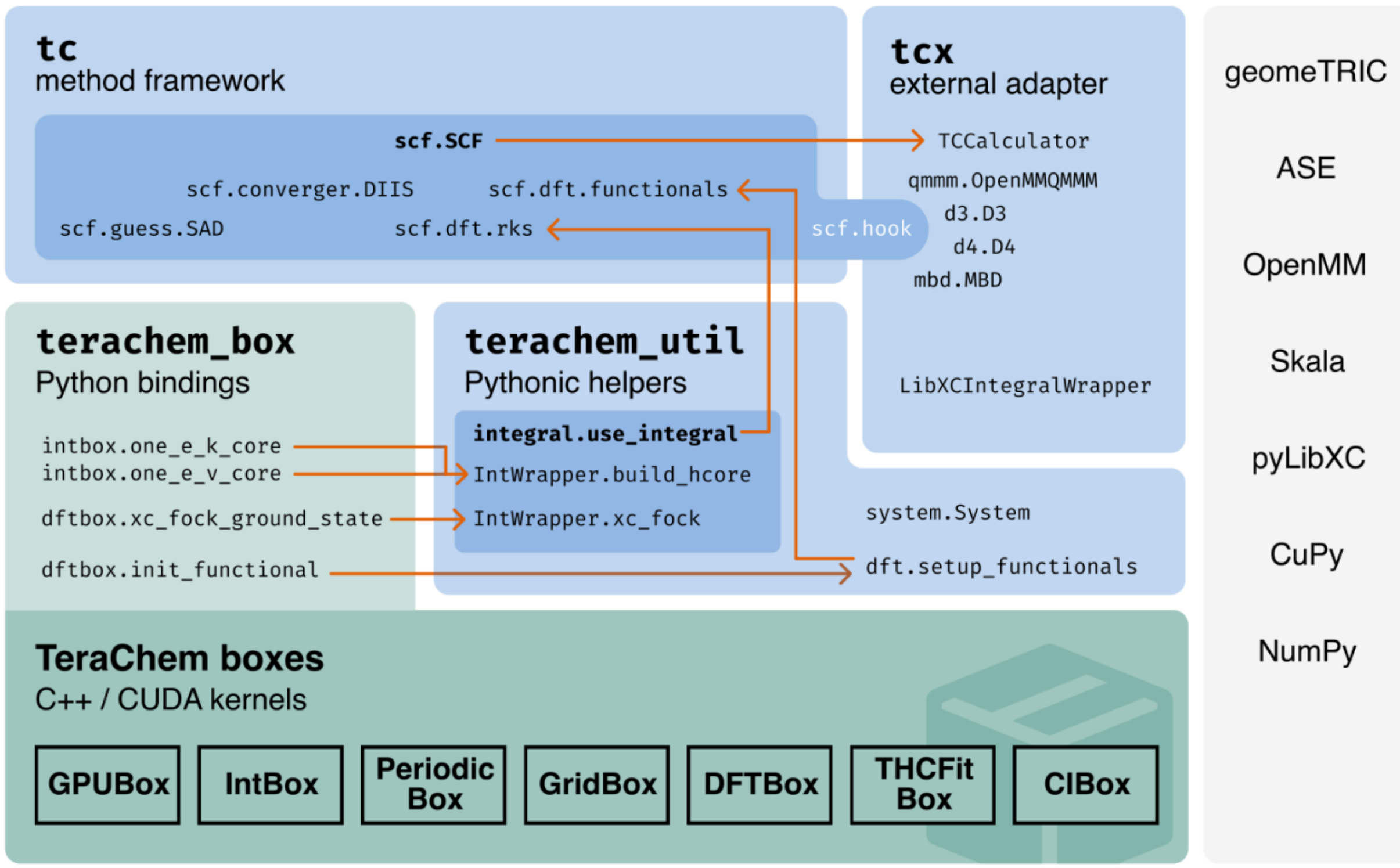

**Figure 1.** Architecture overview of `libterachem.py`. Each layer shows example functions and modules that participate in a restricted DFT calculation. Orange lines represent composition relationships across layers.

### *A. The low-level interface*

TeraChem was originally designed for NVIDIA GPUs and the CUDA platform and this remains the primary target architecture.[23] Its functionality is partitioned into "boxes," self-contained libraries that each provide a reusable family of operations through a stable interface, independent of the calling language.[23] Underlying every box is GPUBox, which owns the GPUs and the host (CPU) and device (GPU) memory that the other boxes draw on. Each box manages its own workspace within this pool. It reserves the memory when it starts, reuses it for successive operations, and releases it when its work is complete. Because GPUBox provides the device underneath, a user of `libterachem.py` works in terms of electronic-structure operations and never manages GPU memory directly.

Once initialized for a calculation, a box holds that calculation's state, such as its loaded basis and current molecular geometry. We call this initialized state a context. A box keeps only one context, a single copy shared across the whole run of the program (one process), and a box that is already running cannot be started a second time. We adopted this restriction deliberately, since confining a box to one context spares it from tracking several concurrent states and simplifies its internal bookkeeping. Parallelism is then handled inside the box, across CPU threads and GPU streams, rather than by the caller. Running several independent contexts requires running as many processes, and coordinating those processes is the province of dedicated workflow packages such as BigChem and ChemCloud.[27]

The current version of `libterachem.py` makes six boxes available from Python. IntBox evaluates the one- and two-electron Gaussian integrals and builds the J and K matrices, together with their gradients, effective core potentials, and density-fitted variants. GridBox constructs the numerical integration grids, computes the Becke partition weights, and evaluates the electron density and its derivatives at the grid points. DFTBox turns these grid quantities into the XC energy, potential, and gradient, using either TeraChem's native functionals or LibXC. THCFitBox builds a low-rank THC representation of the electron-repulsion integrals, forming the collocation matrix $X$ and the central matrix $Z$ from a pruned real-space grid. CIBox is a GPU configuration-interaction solver, providing the diagonalization of the configuration-interaction Hamiltonian, its application to trial vectors, and the reduced density matrices. PeriodicBox provides the integral and XC operations under periodic boundary conditions, including the periodic J and K builds and their gradients. These boxes cooperate within a calculation. In a molecular Kohn–Sham step, IntBox builds the J and K matrices, GridBox evaluates the density on the quadrature grid, and DFTBox turns that density into the XC contribution. Together, these form the Fock matrix. THCFitBox and CIBox instead anchor distinct workflows, the former feeding reduced-scaling correlation methods and the latter solving the configuration-interaction problem from integrals prepared by IntBox. Of these, IntBox and CIBox were described in the previous article on the boxes,[23] while GridBox, DFTBox, THCFitBox, and PeriodicBox postdate it. The THC methodology implemented by THCFitBox is described in Section III-A.

Most of the Python bindings for these boxes (the wrapper code that makes the box operations callable from Python) are automatically machine-generated. Each box declares its operations in a C header, even though the operations themselves are written in C++ and CUDA. We keep the interface in C by design, since a C interface is simple and stable to link against and any language that can call C can in turn call the boxes. Python is one such consumer, and the same headers could serve others. These headers serve as the single source of truth for the binding generation process: the generator reads them and produces both the callable Python wrappers and the companion files that document each function and its argument types, so the two stay in sync as the boxes evolve. The generator parses the headers with libclang and emits the bindings with nanobind.[43-44] Figure 2 outlines this process.

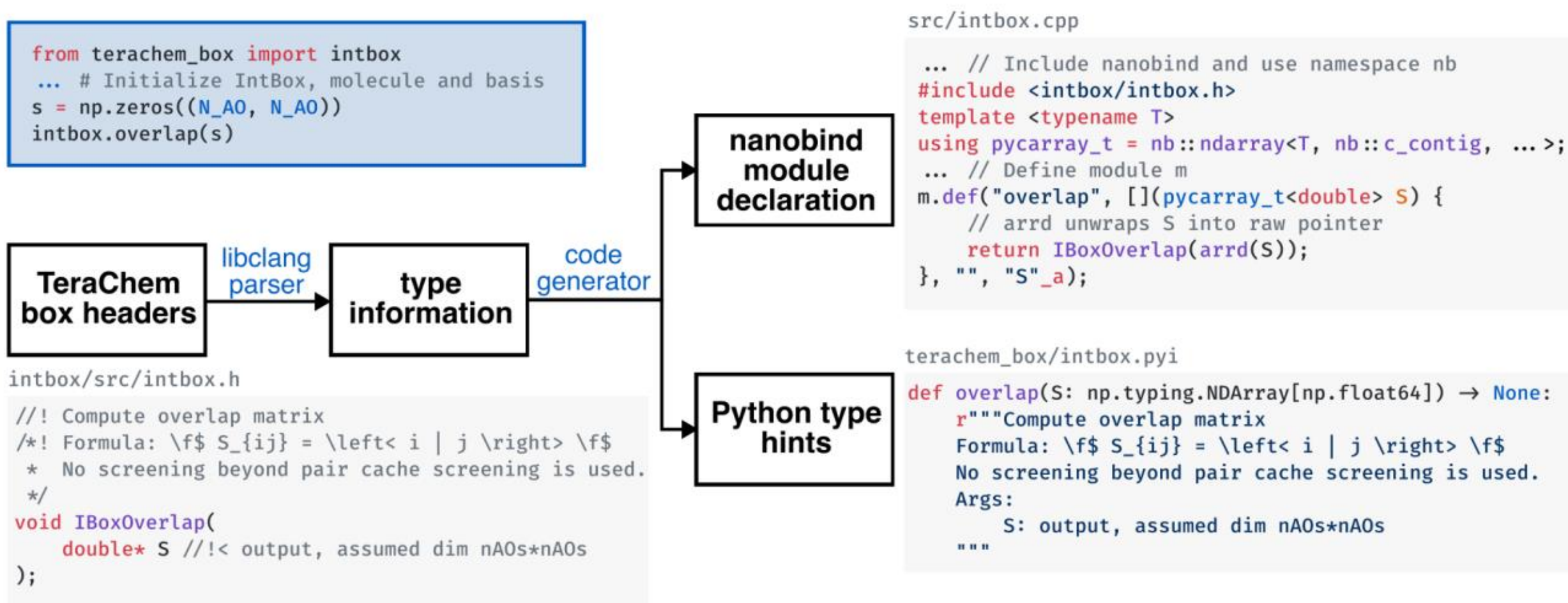

**Figure 2.** Automated generation of the low-level Python interface. The generator parses the boxes' C headers with libclang and produces nanobind wrappers together with synchronized Python documentation and type information.

Arrays are passed between Python and the boxes through host memory. An array is passed directly when its data are already laid out consecutively in the expected numerical type, and is otherwise automatically copied and/or converted. Each box performs its own host–device transfers internally, so device memory is never exposed to Python. CuPy can still be used for Python-side array computations between box calls (e.g., a GPU eigensolve), but its results are copied to the host before the next box call.

On their own, the generated bindings expose each box as a set of low-level routines that carry an internal state and are shared across the running program. The necessary routines for setup and bookkeeping are supplied by `terachem_util`. Examples include the molecular-grid and functional setup that GridBox and DFTBox require, the active-space integral transforms that feed CIBox, and the parsing of standard basis-set definitions into the Gaussian shells the boxes operate on. On top of these box-specific helpers, `terachem_util` adds the shared abstractions needed to assemble an electronic structure method. A `System` object and its periodic counterpart hold the molecular geometry, read from an XYZ file or a coordinate string, and record the charge, spin multiplicity, and, for periodic systems, the cell vectors. A lifetime manager starts each box before use and releases it afterward, enforcing the single active context per box. `terachem_util` also provides wrappers for numerical linear algebra, such as orthogonalization and diagonalization, offloading large matrices to the GPU through CuPy.

One abstraction in `terachem_util` deserves separate mention, because it links the operations to the methods. A method does not depend on the specific form of the underlying Hamiltonian operations, but only asks for specific physical quantities: the core Hamiltonian and overlap together with the J, K, and XC contributions and their gradients. A unified integral interface exposes exactly these, and any implementation that provides them can satisfy the method's requirements. The implementation behind the interface can therefore change while the method that consumes it stays the same. This separation has a precedent inside TeraChem, whose SQMBox already ran the shared method routines over an integral wrapper so that different Hamiltonian operations could feed the same procedures.[24] The integral interface carries this separation into Python, where adapters can supply individual operations or combine implementations to provide the quantities a method requires. `tc` already relies on this interface internally. For example, the self-consistent-field (SCF) methods request the K contribution in the same way whether it is evaluated exactly, by density fitting, by THC, or under periodic boundary conditions. The density-fitted and THC implementations replace only the J and K builds and reuse the remaining operations of the conventional implementation, whereas the periodic implementation supplies all of them. External libraries can supply operations through Python adapters without adding library-specific dispatch to compiled TeraChem or changing the consuming method routines. We use this path in `tcx` (Section II-C), where PyLibXC supplies the XC evaluation and cuEST supplies a set of integral operations through the same interface.

### *B. The method framework:* `tc`

`tc` assembles the low-level operations into complete electronic structure methods. We describe its organization for the SCF methods. A similar organization carries over to more elaborate methods such as complete active space self-consistent-field (CASSCF).

Every SCF method follows the same procedure, regardless of the physics it describes or the convergence accelerator it uses. It builds the core Hamiltonian and overlap, forms an initial density, and enters the iterative process: repeating the Fock build, the molecular orbital optimization, and the density update until the iteration converges. `tc` defines this procedure once, and within it three components are independently replaceable. The *method* sets the physics, namely how the J, K, and XC contributions form the Fock matrix and how the density is rebuilt from the orbitals. `tc` provides restricted, unrestricted, and restricted open-shell references for both Hartree–Fock and Kohn–Sham theory. The *initial guess* seeds the starting density, from the core Hamiltonian or from a superposition of atomic densities. The *convergence accelerator* drives the iteration to self-consistency, either by DIIS or by a second-order orbital-rotation scheme. Each component obeys the interface for its role, so new methods, initial guesses, and convergence accelerators can be added without editing the framework.

The three components are the features that every SCF calculation needs. A method must also make room for features it cannot anticipate, ones that act at several stages of a calculation. Interactive programs such as videogames face the same challenge. A game cannot predict what a player will do, so it runs an event loop, where each interaction triggers the execution of functions registered for it. `tc` organizes an SCF calculation the same way. It exposes a set of *lifecycle hooks*, predetermined points in the calculation at which an added function is invoked. A hook may be triggered before the SCF loop, after each SCF iteration, or after the nuclear gradient evaluation. A function attached there may observe the calculation or adjust it, adding to the core Hamiltonian and the energy before the loop or to the gradient afterward. Treating unforeseen extensions in the same way a videogame treats user interaction may seem like overkill for an electronic structure program, but it cleanly separates the method from its extensions: an extension acts only at the stages it needs, and the core procedure runs unchanged whether or not one is present. As a simple example of how this can work, we describe an implementation of dispersion corrections (Section III-D) and QM/MM (Section IV-B) based on lifecycle hooks.

Every replaceable component has a default and no hooks are required, so none of this machinery needs to be specified in routine use. A user builds a `System`, chooses a method, and runs, optionally replacing the convergence accelerator or the initial guess. Figure 3 contrasts the two levels on the same water BLYP calculation, with identical basis, thresholds, and convergence criterion, yielding the same final energy. At the low level, we write the Fock build, a CuPy eigensolve, the density update, and the SCF loop by hand. At the higher level (with `tc`), we choose only the components and `tc` orchestrates the iterations, so the same calculation takes only a few lines of code. A user can therefore work at whichever level a task calls for, dropping to the operation level (`terachem_box`/`terachem_util`) to prototype a method and rising to `tc` to run an established one.

**(a) Low-level interface**

```python
def fock2density(fock, x):
    fockp = x.T @ fock @ x
    fockp_dev = cupy.asarray(fockp)
    _, cp_dev = cupy.linalg.eigh(fockp_dev)
    cp = cupy.asnumpy(cp_dev)
    c_occ = c(x @ cp)[:, : mol.n_electrons // 2]
    return (c_occ @ c_occ.T) * 2

mol = terachem_util.system.System.from_xyzfile("h2o.xyz")
maxiter = 100
dmax_conv = 1.0e-6
damp = 0.5
basis = "def2-svp"
functionals = [("slaterx", 1.0), ("beckex", 1.0), ("lypc", 1.0)]

use_intw = terachem_util.integral.use_integral(mol, basis, dft=True)
with use_intw as intw:
    terachem_util.dft.setup_functionals(*functionals)
    e_nuc = intw.nn_repulsion()
    hcore = intw.build_hcore()
    s, x = intw.compute_overlap()
    den = fock2density(hcore, x)  # H core guess
    energy = 0.0
    for step in range(1, maxiter + 1):
        fock = np.copy(hcore)
        intw.j_fock(den, fock)
        energy = np.einsum("pq,pq->", fock + hcore, den) / 2 + e_nuc
        e_xc, _ = intw.xc_fock(den * 0.5, fock)
        energy += e_xc
        error = x.T @ (fock @ den @ s - s @ den @ fock) @ x
        dmax = np.max(np.abs(error))
        if dmax < dmax_conv:
            break
        den = damp * den + (1 - damp) * fock2density(fock, x)
```

**(b) Method framework (`tc`)**

```python
mol = terachem_util.system.System.from_xyzfile("h2o.xyz")
scf = tc.scf.SCF(
    mol,
    basis="def2-svp",
    method=tc.scf.dft.rks(
        tc.scf.dft.functional("blyp")
    ),
)

with scf.session() as s:
    energy = s.energy()
```

**Figure 3.** Two implementations of SCF at two abstraction levels of `libterachem.py`, illustrated schematically for the same BLYP/def2-SVP calculation on $H_2O$: (a) the low-level interface and (b) the `tc` method framework. Shaded regions indicate corresponding execution code. Imports are abridged for clarity.

### *C. External library integration: `tcx`*

`tcx` connects the method framework and the operations beneath it to specific external tools. It adapts each external library to an interface defined by tc or `terachem_util`, as appropriate. We distinguish three cases.

An external library may supply one or more of the Hamiltonian operations a method requests. `tcx` then routes it through the integral interface. PyLibXC[40] is one example, supplying the XC evaluation and reporting the hybrid and range-separation parameters needed by the Fock build. GridBox and IntBox take care of the grid, J, and K operations. NVIDIA's cuEST, introduced in Section I, is a complementary case.[32] Because `tcx standardizes` the interface, it is straightforward to have cuEST supplying the overlap, core Hamiltonian and density-fitted J and K builds in place of IntBox.

An external library may also act around certain stages of a complete method. `tcx` attaches these through the framework's lifecycle hooks, each entering at its natural stage. A dispersion correction adds an energy and gradient after the electronic calculation and never enters the Fock matrix. In contrast, the electrostatic embedding of an OpenMM QM/MM calculation must be applied before or within the SCF loop, modifying the core Hamiltonian and therefore changing the SCF wavefunction.[41] Each correction is an independent module.

An external library may finally consume a complete method from the outside. `tcx` presents a `tc` method as a calculator for the Atomic Simulation Environment (ASE), a widely used Python framework for atomistic simulation.[42] ASE's geometry optimizers, molecular-dynamics integrators, and analysis tools all obtain the energy and forces of a system through a single `Calculator` interface. By implementing that interface, a `tc` method becomes one more ASE calculator, and any ASE-compatible driver can run complete GPU calculations through it without any knowledge of TeraChem. We use this route to drive velocity Verlet molecular dynamics (Section IV-B) and geometry optimization (Section III-E). Because the same calculator object is reused across successive geometries, it also carries the converged density from one step over as the initial guess for the next, reducing the number of SCF iterations along a trajectory.

In each case the adaptation to a particular library is confined to `tcx`. The method framework and the boxes beneath it need no knowledge of any specific external library.

## III. Interoperable and flexible workflows

`libterachem.py` supports both fine-grained method development and integration with external packages. The THC and automatically differentiated XC examples work directly with box operations and their exposed intermediate quantities, combining them with Python libraries for numerical experimentation. The remaining examples use complete `tc` methods and connect them to external correction and simulation packages through `tcx`.

### *A. Rapid experimentation with tensor hypercontraction*

Tensor hypercontraction (THC) approximates the four-index electron-repulsion integral tensor in the fully separable form

$$(\mu\nu|\lambda\sigma) \approx \sum_{PQ} X_\mu^P X_\nu^P Z^{PQ} X_\lambda^Q X_\sigma^Q, \quad (1)$$

where $X_\mu^P$ is a weighted atomic-orbital (AO) collocation factor evaluated at real-space grid point $P$, and $Z^{PQ}$ is the central grid–grid interaction matrix.[35, 45] By separating all four AO indices, THC can reduce the storage and contraction costs of the electron-repulsion integral tensor and has enabled reduced-scaling formulations of MP2,[45] SOS-MP2,[36] CASPT2,[46] and coupled-cluster methods.[47]

The GPU-accelerated THCBox library and its applications in correlated methods were described in the previous TeraChem software article.[23] The developments considered here instead use THCFitBox, a redesigned infrastructure for constructing least-squares THC representations fitted to TeraChem's density-fitting integral backend. Through `libterachem.py`, individual operations for molecular-grid generation, grid pruning, AO collocation, and THC-factor construction can be invoked independently from Python. Production GPU kernels can therefore be combined with parameter scans and numerical analysis of intermediate quantities, rather than being reachable only through completed energies and gradients.

The real-space grid is a central algorithmic choice in grid-based THC. An undersized grid may not adequately span the AO pair-product space, whereas an unnecessarily dense grid increases the dimensions of $X$ and $Z$ and introduces near-linear dependencies into the least-squares fit. Pivoted-Cholesky pruning provides a systematic means of removing this redundancy while retaining the dominant contributions of the pair-product space.

For the least-squares variant of THC, the overlap between the implicit pair-product vectors associated with grid points $P$ and $Q$ defines the grid-pair metric

$$S^{PQ} = \sum_{\mu\nu} X_\mu^P X_\nu^P X_\mu^Q X_\nu^Q = \left(\sum_\mu X_\mu^P X_\mu^Q\right)^2. \quad (2)$$

Redundant grid points lower the rank of $S$, leading to a large condition number. Because the dimensions of both $S$ and $Z$ are determined by the number of retained points, this redundancy also increases the cost of THC construction and downstream contractions.

Pivoted-Cholesky pruning selects grid points greedily according to their residual contribution to $S$ and terminates when the largest remaining contribution falls below a prescribed threshold. Building on its previous use as a grid-selection criterion for least-squares THC,[48] THCFitBox provides an efficient implementation for dense parent grids.

Figure 4 illustrates this fitting-level access through a simple parameter scan. Following basis and molecular-grid setup, the Cholesky pruning itself is performed by the GPU-accelerated THCFitBox implementation, while Python is used to vary the pruning threshold, access the resulting AO collocation matrix $X$, and evaluate the condition number $\kappa(S) = \|S^{-1}\|\|S\|$ with NumPy.[49] The highlighted calls in the code block (Figure 4, left) identify the THCFitBox operations used to prune the grid and construct $X$. At tight thresholds, approximately 350 points per atom are retained and a large $\kappa(S)$ indicates severe ill-conditioning. Increasing the threshold removes near-dependent directions, reducing both the retained grid size and the condition number.

Conditioning alone does not determine an optimal threshold, since overly aggressive pruning may also remove information required for accurate integrals or energies. A complete pruning study would therefore supplement this diagnostic with recovered-integral and energy benchmarks. For the present scan, `libterachem.py` keeps the calculation state initialized within a single Python workflow, so each threshold can be tested without generating a new TeraChem input, launching a complete calculation, or parsing its output. Intermediate quantities are accessed directly, and only the data required for the analysis need be materialized.

**THCFitBox**

```python
from terachem_box import thcfitbox
# After basis and molecular-grid setup
for eps in EPS_VALUES:
    params = THCFitBoxParams()
    params.set_defaults()
    params.GridPruneMode = THCFitBoxGridPruneMode.CHOLESKY
    params.GridPruneCholeskyEps = eps

    npt = thcfitbox.init_and_prune(grid.n_points, grid.point_xyz, ..., params)
    x = np.zeros((npt, nao), dtype=np.float64)
    thcfitbox.build_x(x)

    cond = np.linalg.cond((x @ x.T) ** 2)
```

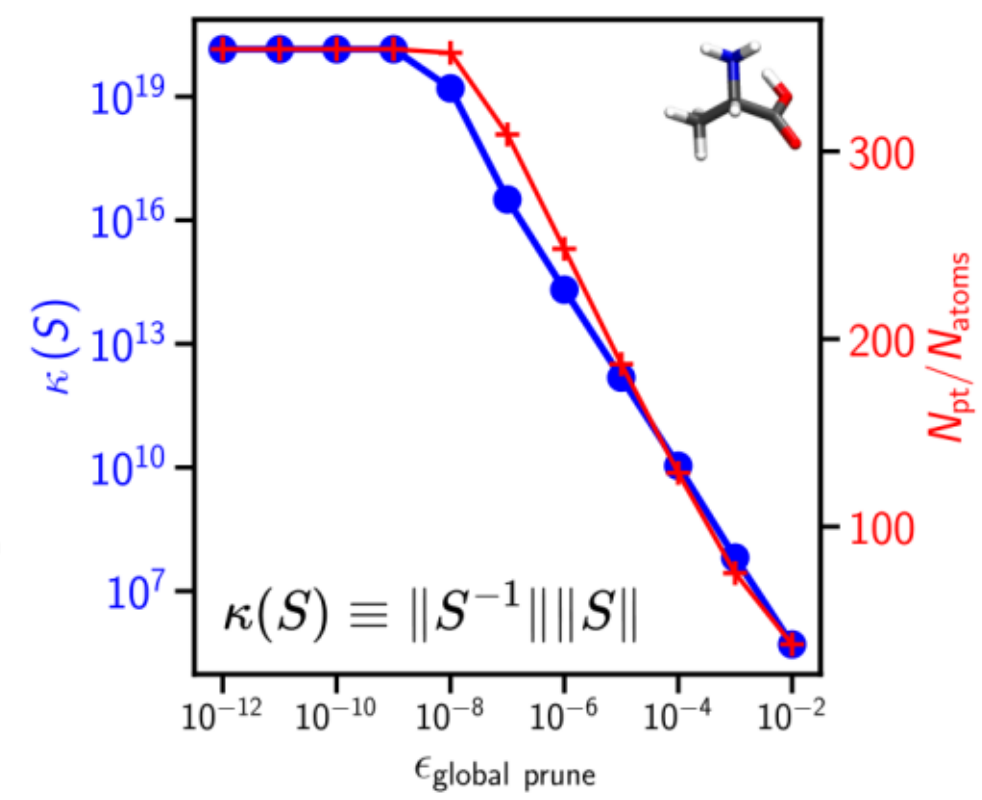

**Figure 4.** Analysis of Cholesky grid pruning for least-squares tensor hypercontraction on alanine with cc-pVDZ/cc-pVDZ-RIFIT.[50-51] Left: abbreviated `libterachem.py` workflow following basis and molecular-grid setup. Highlighted calls denote THCFitBox operations that perform global Cholesky pruning over a range of thresholds and construct the corresponding AO collocation matrix $X$. NumPy is then used to evaluate the condition number of the resulting grid-pair metric $S$. Right: condition number $\kappa(S)$ (blue circles, left axis) and retained grid points per atom (red crosses, right axis) as functions of the pruning threshold.

### *B. Extensible XC evaluation*

In numerical Kohn–Sham density functional theory, molecular-grid construction, density collocation, XC evaluation, and contraction of the resulting XC potential into the Fock matrix are naturally separable operations.[52-53] TeraChem provides optimized native implementations of a focused set of commonly used XC functionals, together with GPU-accelerated grid, J, and K.

*1. Conventional functionals through PyLibXC*

LibXC[40] provides standardized implementations of a broad range of local, semilocal, hybrid, and range-separated hybrid XC functionals. A compiled C/C++ adapter to LibXC was developed after the previous TeraChem software article and has since been used internally, extending TeraChem's functional repertoire to include meta-GGA and hybrid meta-GGA forms. `libterachem.py` also provides a Python-level LibXC interface as part of `tcx`.

Figure 5 illustrates a range-separated hybrid calculation using the CAM-PBEH functional.[54] At the method level, a LibXC integral wrapper supplies the requested functional to the DFT method and is added as an integral layer to the SCF calculation. The lower portion of the figure expands the operations performed by this interface. For each SCF iteration, GridBox evaluates the electron density and its Cartesian gradient on the molecular quadrature grid. PyLibXC is supplied with the density and the gradient invariant $\sigma = |\nabla\rho|^2$, and returns the XC energy per particle together with its derivatives with respect to $\rho$ and $\sigma$. GridBox integrates $\rho\epsilon_{\mathrm{xc}}$ to obtain the XC energy and contracts the corresponding local potential into the Fock matrix. IntBox separately evaluates the J and range-separated K contributions. The CAM coefficients reported by LibXC are translated to the full-range and error-function-scaled convention expected by the IntBox K kernel before the Fock build.

The present calculation uses the CPU implementation of PyLibXC and is intended to illustrate functional interoperability rather than benchmark end-to-end performance. Since PyLibXC is installed independently of TeraChem, newly added LibXC functionals can be accessed through `libterachem.py` as soon as they become available, without modifying or recompiling the TeraChem codebase. The converged energy was cross-checked against both the compiled TeraChem–LibXC interface and an independent PySCF[10] calculation using the same LibXC functional and Cartesian Gaussian basis, with microhartree agreement.

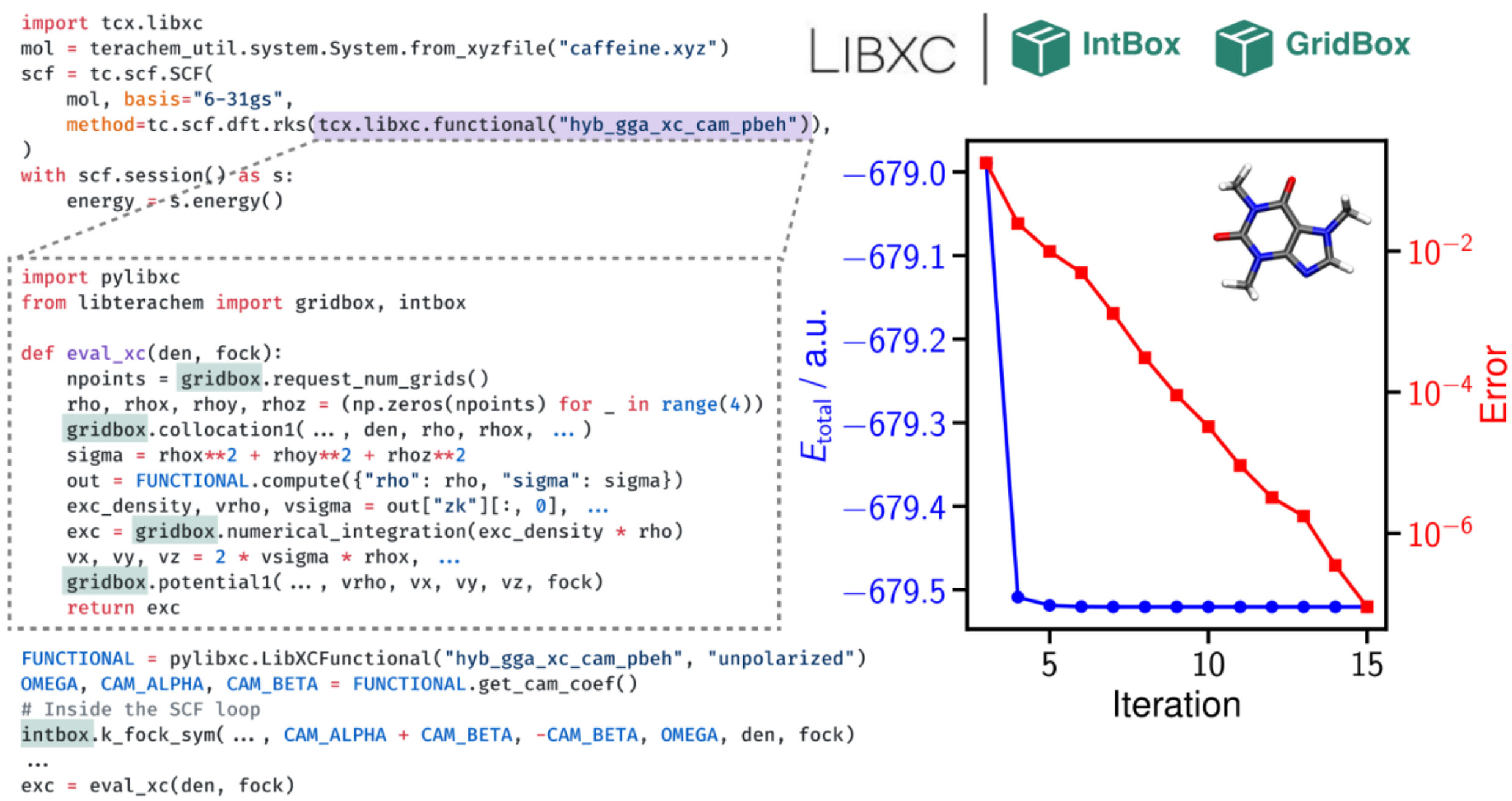

**Figure 5.** XC evaluation using PyLibXC for caffeine with the 6-31G* basis.[55-56] Upper left: the user-facing CAM-PBEH SCF configuration, where the LibXC integral wrapper supplies the requested functional to the DFT method. Lower left: an expanded view of the underlying XC implementation, where PyLibXC supplies the XC energy and functional derivatives, GridBox evaluates the density and its gradient and contracts the resulting XC potential into the Fock matrix, and IntBox evaluates the range-separated K contribution. Right: convergence of the total energy and the SCF error, defined as the maximum element of the commutator residual.

The same decomposition can accommodate other Python-accessible XC models, provided that their required density features and quadrature conventions are exposed through a well-defined interface.

*2. Differentiable machine-learned functionals with Skala*

Whereas LibXC exposes predefined functional forms and their derivatives through a standardized interface, machine-learned XC models may require model-specific feature construction, with the corresponding potential obtained by automatic differentiation. Skala[18] is a message-passing neural-network XC functional distributed with a reference PyTorch[57] implementation and also integrated with GauXC[14] and GPU4PySCF.[12] Its model evaluation is independent of the host electronic-structure program. We assemble its input features directly from TeraChem's exposed numerical operations, keeping both the model and feature construction modifiable from Python.

Skala combines spin-resolved densities, density gradients, and kinetic-energy densities with geometric and atomic-quadrature information used by its nonlocal layers. At each SCF iteration, GridBox evaluates the density-derived fields, the PyTorch model computes the integrated

XC energy, and automatic differentiation supplies the derivatives needed for GridBox to assemble the XC contribution to the Fock matrix (Figure 6, left). A Python adapter arranges the collocated fields into contiguous atomic grids for Skala and restores TeraChem's spatial-box ordering for potential assembly. Because automatic differentiation acts on the integrated energy, the adapter uses unit contraction weights to avoid applying quadrature weights a second time, while the physical weights remain part of the model evaluation. The point permutation is cached, the model is loaded once, and Skala and TeraChem run in the same process on one GPU.

TeraChem applies a Becke molecular partition[52] to the atomic quadrature weights and removes points with negligible partitioned weights. Although those points contribute little to molecular quadrature, they can still contribute to Skala's nonlocal atomic integrations, which use unpartitioned weights. To reproduce the reference integration conventions, we therefore apply the reference implementation's Treutler-adjusted Becke partition[58] and retain the complete atomic quadrature sets.

Figure 6 (right) compares the grid convergence of Skala with that of TPSS[59] and $r^2$SCAN.[60] The calculations use the same Python-driven SCF procedure and GridBox operations, with functional energies and derivatives supplied by PyTorch and PyLibXC, respectively. Across TeraChem grid types 1–5, Skala's converged total energy approaches the PySCF level-6 reference nonmonotonically, reaching an absolute deviation of 0.008 kcal mol$^{-1}$ at type 5. On a single NVIDIA A100, the mean SCF iteration time, excluding the first iteration, increases from 0.29 to 4.36 s between grid types 1 and 5. This example demonstrates incorporation of an external differentiable functional and control over its numerical integration within a Python-driven SCF calculation, while retaining TeraChem's GPU operations and requiring no model-specific XC-potential implementation inside TeraChem.

```python
# Initialize TeraChem before importing Skala/PyTorch
from terachem_box import gridbox
g = GridFeatureBuilder(grid, thresp, thredp)  # gridbox.init_molecular_grid w/ uniform weights,
                                              # plus the point permutation
import skala.functional
import torch
dev = torch.device("cuda")
functional = skala.functional.load_functional("skala-1.1", device=dev)
leaf = lambda a: torch.tensor(a, dtype=torch.float64, device=dev, requires_grad=True)
# The grid features are fixed for the whole SCF:
#   grid_weights, atomic_grid_weights,
#   grid_coords, atomic_grid_sizes, coarse_0_atomic_coords
static = {k: torch.as_tensor(v, device=dev) for k, v in g.static.items()}

# Once per SCF iteration
rho, rhox, ..., lap, tau = (np.zeros(grid.n_points) for _ in range(6))
gridbox.collocation2(thresp, thredp, den_alpha, rho, rhox, ..., lap, tau)
grad = np.stack([rhox, rhoy, rhoz])
rho, tau, grad = (leaf(g.order_tc_to_skala(t)) for t in (rho, tau, grad))

exc = functional.get_exc({
    **static,
    "density": torch.stack([rho, rho]), # closed shell: one leaf,
    "grad": torch.stack([grad, grad]),  # shared by both spins
    "kin": torch.stack([tau, tau]),
})

d = torch.autograd.grad(exc, (rho, grad, tau)) # Vxc, by autodiff
d_rho, d_grad, d_tau = (g.order_skala_to_tc(t.cpu().numpy()) for t in d)
gridbox.potential2(thresp, thredp, (0.5 * d_rho), *(0.5 * d_grad),
                   np.zeros_like(d_rho), (0.5 * d_tau), fock)
                   # chain rule from D_alpha = D_beta = D/2
```

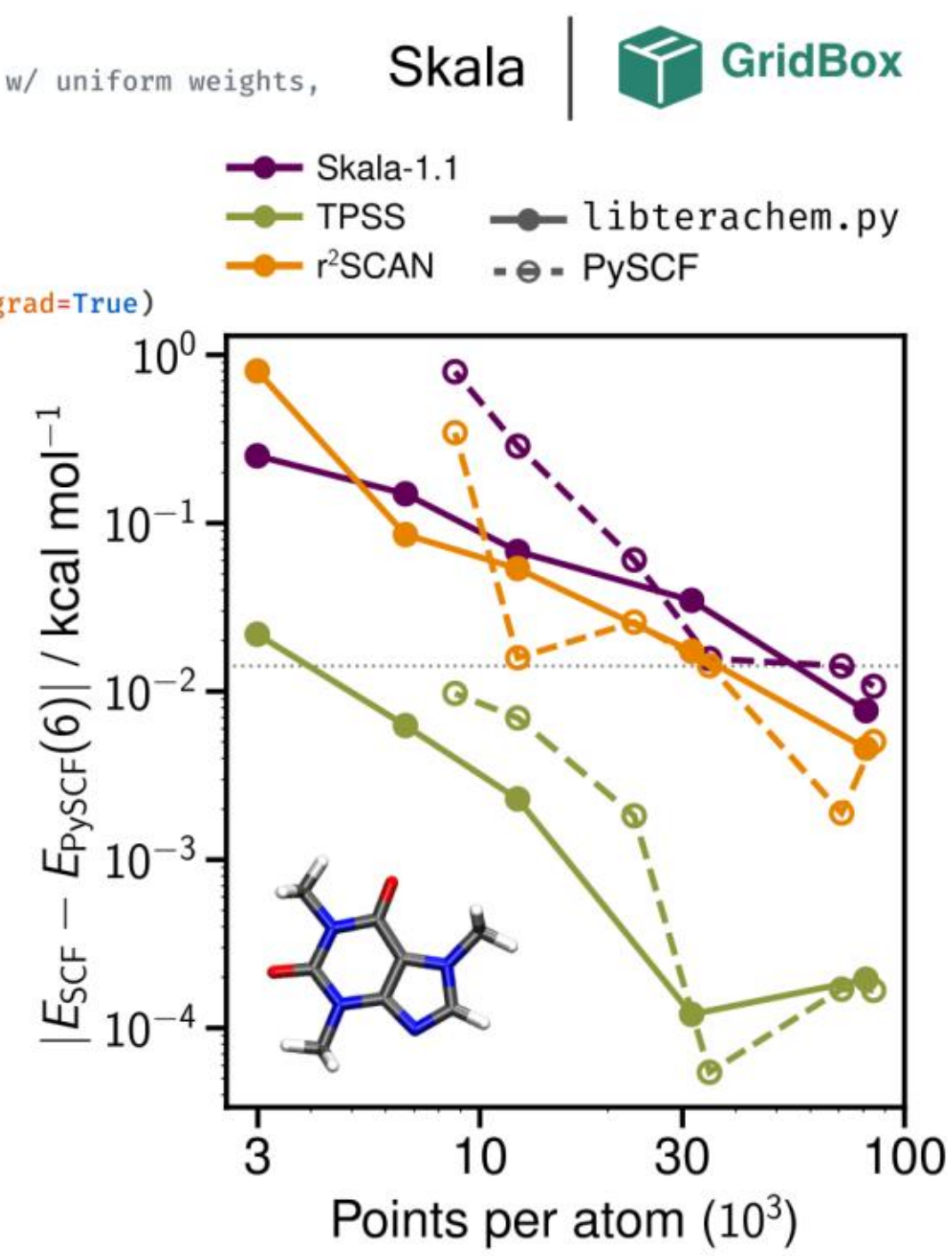

**Figure 6.** Skala integration and grid convergence for caffeine with the 6-31G* basis. Left: abbreviated XC workflow with directly invoked GridBox operations highlighted. Right: absolute deviations of converged total energies from each functional's PySCF grid-level-6 reference. Solid lines show `libterachem.py` on complete TeraChem grids of types 1–5, and dashed lines show PySCF on its own grids. The dotted line marks the largest Skala reference-energy change at PySCF levels 7 and 8 relative to level 6 (0.014 kcal $mol^{-1}$).

### *C. Periodic systems*

Extended systems can also be treated under periodic boundary conditions using PeriodicBox, which implements GPU-accelerated Gaussian-orbital integrals within the Γ-point approximation.[39] PeriodicBox is used in the same way as IntBox, except that a periodic system is instantiated with a unit cell in addition to the atomic positions. The integral setup selects PeriodicBox based on the periodic system type.

Figure 7 shows an equation-of-state calculation for diamond using `tc` and `tcx`, in which the volume of the eight-atom conventional cubic cell is scanned and the energies are fitted to the Birch–Murnaghan equation.[61] ASE's bulk builder constructs the unscaled cell and the volume scan, a `tc` SCF calculation gives the energy at each volume, and ASE performs the Birch–Murnaghan fit.

```python
from ase.build import bulk
from ase.eos import EquationOfState
import tcx.calculator

HARTREE_TO_EV = 27.2114
diamond = bulk("C", "diamond", a=3.56679, cubic=True)
volumes, energies = [], []

for scale in np.linspace(0.95, 1.15, 14):
    atoms = diamond.copy()
    atoms.set_cell(diamond.cell * scale, scale_atoms=True)
    scf = tc.scf.SCF(
        tcx.calculator.from_ase_atoms(atoms),
        basis="6-31g",
        method=tc.scf.dft.rks(tc.scf.dft.functional("blyp"))
    )
    with scf.session() as session:
        energy = session.energy()
    volumes.append(atoms.get_volume())
    energies.append(HARTREE_TO_EV * energy)

eos = EquationOfState(volumes, energies, "birchmurnaghan")
v0, e0, B0 = eos.fit()
```

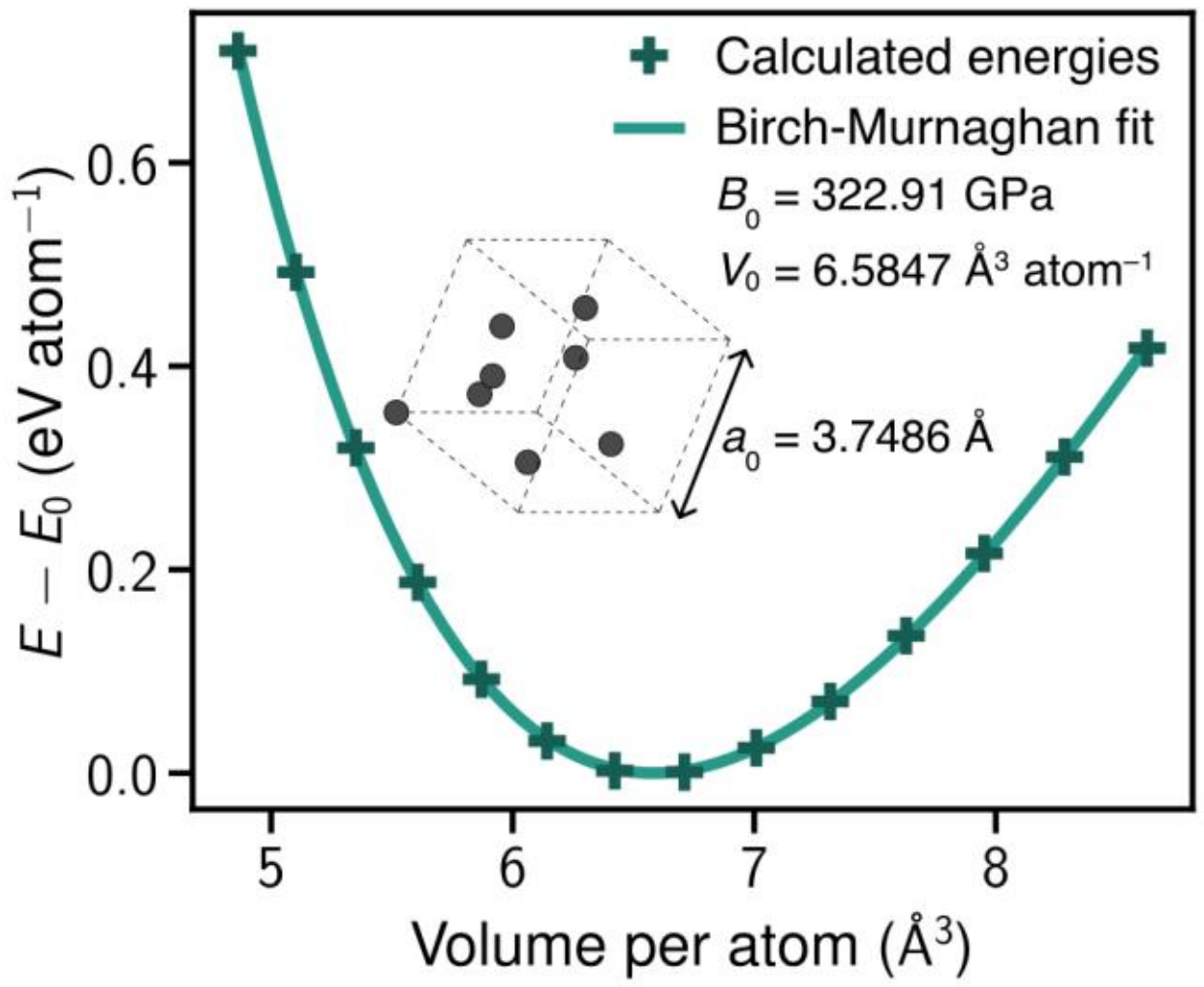

**Figure 7.** Periodic equation-of-state calculation for diamond carbon at the BLYP/6-31G level. Left: ASE constructs and isotropically scales the eight-atom conventional cell, `tcx converts each structure into a PeriodicSystem,` and `tc` evaluates its energy before ASE performs the Birch–Murnaghan fit. Right: calculated energies per atom and the fitted curve, with the inset showing the cell geometry and equilibrium lattice parameter. Energies are referenced to the fitted minimum, $E_0/N = -1034.62$ eV atom$^{-1}$.

With BLYP/6-31G, this example gives a lattice constant for diamond of 3.748 Å, 5% larger than the experimental value of 3.56679 Å. Much of this discrepancy is expected, since the calculation uses a small basis set and the Γ-point approximation, which requires a large supercell to approach the thermodynamic limit.

Periodic SCF accepts the same components and lifecycle hooks as molecular SCF, including the dispersion corrections of the next subsection.

### *D. Composable dispersion corrections*

Dispersion-dominated interactions lie beyond the reach of common semilocal and hybrid functionals, which recover little of the long-range correlation that binds stacked and van der Waals complexes.[62] A dispersion correction restores this contribution as an additive term on the electronic energy and gradient. Several such corrections are in wide use, from the D3[63-64] and D4[65] models to many-body dispersion (MBD).[66-67] They differ in physical content and cost, and no single model is definitive, so comparing them on a given system is a routine step in method selection. As introduced in Section II, `tcx` attaches each correction to a DFT calculation through the framework's lifecycle hooks, contributing an energy and gradient around the electronic calculation without entering the Fock matrix. Because the three corrections obey the same interface, they are interchangeable.

We illustrate this on the binding curve of the cofacial coronene dimer, a stacked aromatic pair held together almost entirely by dispersion. Both monomers are kept rigid and displaced symmetrically along the stacking axis, which scans the interplanar separation from 2.8 to 8.0 Å. The interaction energy is defined as $E_{\text{int}} = E_{\text{dimer}} - 2E_{\text{monomer}}$. Every point uses the same PBE0/def2-SVP calculation with identical grid and convergence settings, and because dispersion acts after the electronic calculation, the three curves differ only in the dispersion model. We use D3 with Becke–Johnson damping (D3(BJ)),[64] D4 with its three-body Axilrod–Teller–Muto term,[65] and range-separated self-consistently screened MBD (MBD@rsSCS).[67] We evaluate MBD through libMBD's Python interface.[62]

Figure 8 shows the result. The upper half of the code block on the left is the complete method definition. The only line that changes between the three corrections is the highlighted line, which selects the dispersion component. On the right, bare PBE0 shows no appreciable binding on this scale, with a shallow minimum below 0.1 kcal/mol, so the binding is almost entirely due to dispersion. All three dispersion corrections predict essentially the same optimal separation at 3.6 Å. The well depths are spread over ≈ 2≈ kcal/mol, with D4 (−19.0 kcal/mol) being more strongly binding than D3 (−18.4 kcal/mol) which is more strongly binding than MBD (−17.0 kcal/mol). The dispersion-corrected energy curves are all nearly parallel over this scan.

Every point requires the same GPU electronic-structure work (the two-electron integrals, the SCF iterations, and the XC build on the grid). The dispersion correction is a small addition to this, so exchanging D3 for D4 or MBD touches only that one component and leaves the electronic-structure calculation untouched. The three-model comparison thus requires only one SCF calculation per geometry. A new dispersion model can be included in the same way, as an independent component, without any change to the method framework or the operations beneath it.

```python
import tcx.mbd
mol = terachem_util.system.System.from_xyzfile("coronene-dimer.xyz")
scf = tc.scf.SCF(
    mol, basis="def2-svp",
    method=tc.scf.dft.rks(tc.scf.dft.functional_libxc("pbe0")),
    external_methods=[tcx.mbd.MBD("pbe0")],
)
with scf.session() as s:
    energy = s.energy()
    gradient = s.gradient()
```

```python
...
from terachem_box import gridbox
def _volume_ratios(mol, bi, den, den_free):
    point_xyz, rho = np.empty((npoints, 3)), np.zeros(npoints)
    gridbox.get_grid(point_xyz, np.empty(npoints))
    gridbox.collocation0( ... , den, rho)
    rho_free = np.zeros((len(mol.atm_names), npoints))
    for a, name in enumerate(mol.atm_names):
        cols = np.concatenate(bi.ranges_atom_cart[a])
        d_a = np.zeros((nao, nao))
        d_a[np.ix_(cols, cols)] = den_free[name]
        gridbox.collocation0( ... , d_a, rho_free[a])
    np.clip(rho_free, 0.0, None, out=rho_free)
    rho_promol = np.maximum(rho_free.sum(axis=0), 1e-30)
    ratios = np.empty(len(mol.atm_names))
    for a in range(natoms):
        r3 = np.linalg.norm(point_xyz - mol.geometry[a], axis=1) ** 3
        w = rho_free[a] / rho_promol  # Hirshfeld weight
        v_eff = gridbox.numerical_integration(w * r3 * rho)
        v_free = gridbox.numerical_integration(r3 * rho_free[a])
        ratios[a] = v_eff / v_free
    return ratios
...
```

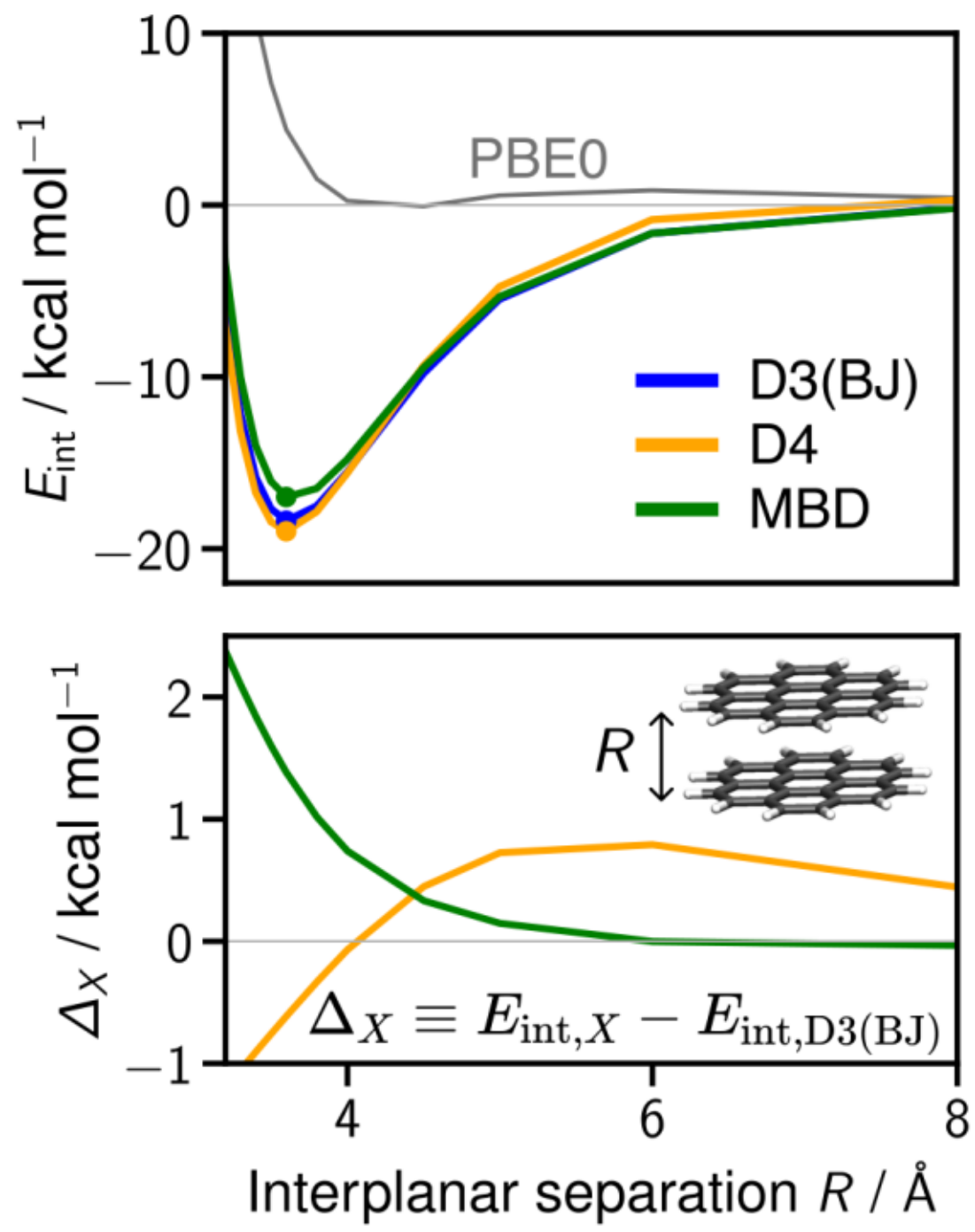

**Figure 8.** Implementation and comparison of composable dispersion corrections. Upper left: the user-facing SCF configuration that attaches MBD through `external_methods`, where the line highlighted in purple may also select other dispersion methods (D3 or D4) and is the only line that differs between the three corrections. Lower left: A snippet of the `mbd` module, where GridBox (highlighted in green) operations construct the Hirshfeld volume ratios required by MBD. Upper right: binding curves for the uncorrected and dispersion-corrected methods. Filled circles mark their lowest sampled points. Lower right: differences relative to the D3(BJ)-corrected curve. The molecular inset indicates the scanned coordinate *R*.

### *E. Geometry optimization through geomeTRIC*

The examples so far evaluate energies at fixed geometries. Many workflows instead require an external optimizer to move the nuclei. Through the ASE calculator interface of Section II, a complete `tc` method can serve as an energy-and-force provider for such workflows. We demonstrate this by passing the PBE0/def2-SVP method with the MBD correction used in the preceding subsection to the geomeTRIC optimizer.[68]

We revisit the coronene dimer, this time relaxing the relative pose of the two monomers at each stacking distance. The fixed scan of the previous subsection held the dimer perfectly cofacial. Here we keep each monomer internally rigid but let geomeTRIC optimize their relative arrangement, namely the lateral displacement, the rotation about the stacking axis, and the tilt, at a fixed centroid separation. geomeTRIC's rigid-body coordinates express this directly. The intramolecular coordinates are constrained, so only the rigid-body degrees of freedom of each monomer remain free, and a distance constraint holds the centroid separation. Repeating the relaxation across a range of separations yields a relaxed binding curve.

Figure 9 contrasts this rigid-body-optimized PBE0+MBD curve with the fixed cofacial MBD curve of Figure 8. The driver code on the left of the figure shows that the composed method is wrapped as an ASE calculator, each monomer is constrained to move as a rigid body, the centroid separation is fixed, and geomeTRIC optimizes the remaining relative pose. Relaxing the eclipsed arrangement lowers the interaction energy relative to the fixed cofacial scan, as the monomers slip away from the perfectly stacked geometry.

The electronic structure method is identical to the one used for the fixed scan, and only the driver changes: geomeTRIC proposes each geometry, the ASE calculator returns the energy and forces, and the composed PBE0+MBD method supplies them without modification.

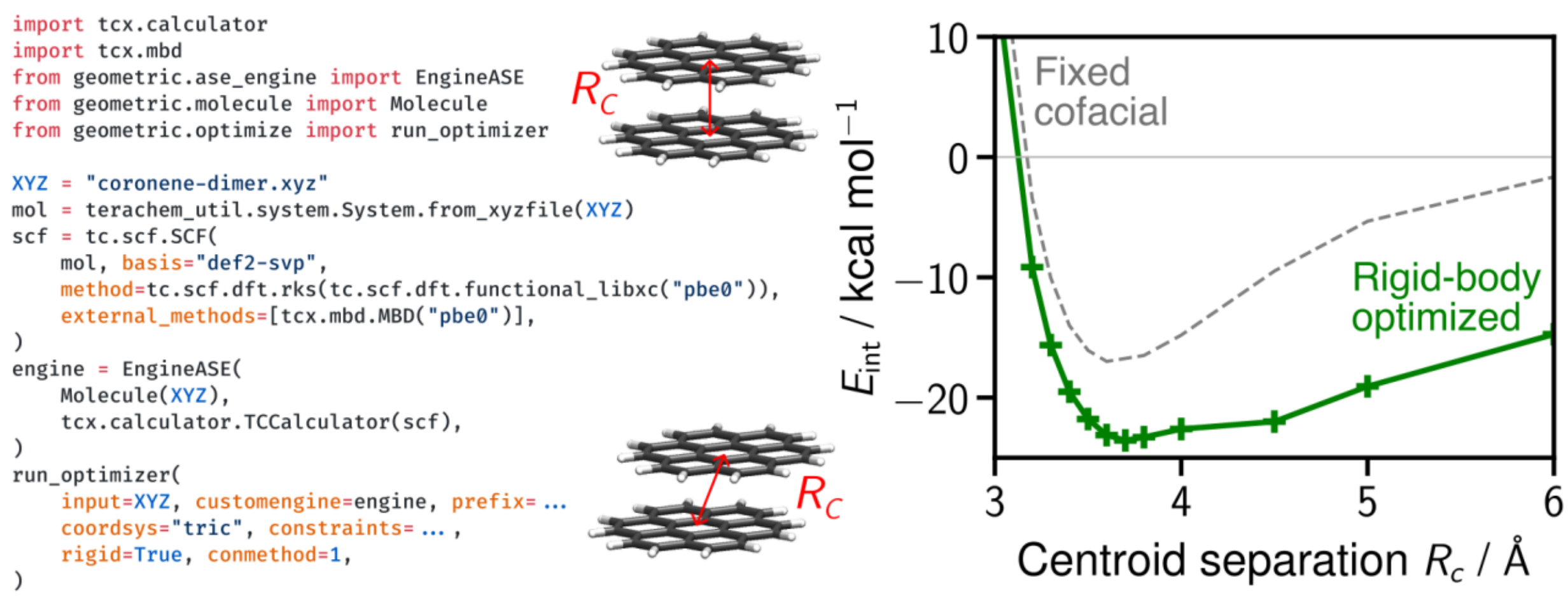

**Figure 9.** Rigid-body optimization of the coronene dimer with geomeTRIC using PBE0/def2-SVP with the MBD correction. Left: The driver connects the composed `tc` method to geomeTRIC through ASE. Right: The resulting interaction-energy curve shows the stabilization obtained by optimizing the relative monomer pose at each fixed centroid separation, $R_C$. The fixed cofacial curve from Figure 8 is shown for comparison. The top and bottom molecular visualizations represent the initial cofacial and rigid-body optimized geometries at $R_C = 3.7$ Å, respectively.

## IV. Performance and Application

We now evaluate the computational performance of the Python interfaces, first by quantifying SCF overhead and multi-GPU strong scaling, then by measuring end-to-end throughput in a protein QM/MM simulation, and finally by characterizing the cost of multireference excited-state calculations on the resulting trajectories.

### *A. SCF overhead and multi-GPU scaling*

Section II described the Python layer as forwarding each operation to the boxes beneath it, and here we quantify the cost of the associated computational overhead. We run identical $\omega$B97X/def2-SVP single points through both `libterachem.py` and the native TeraChem executable, on DNA fragments of one to ten base pairs (655 to 7550 atomic orbitals) on a single NVIDIA H100 PCIe 80 GB GPU. The two programs share every calculation setting, namely the initial guess, integration grid, convergence threshold, and precision. Both rebuild the complete J, K, and XC contributions to the Fock matrix every cycle rather than updating them incrementally, and both converge in the same number of cycles. Both also evaluate the XC term through the same DFTBox engine, so the J, K, and XC builds invoke identical GPU kernels on the two sides. Each timing is a mean over five fresh-process runs, reported with its standard error. GPU startup and kernel compilation are excluded and the first SCF cycle is dropped, so the times are steady-state per-cycle. The converged energies agree to within $8 \times 10^{-8}$ Ha.

On these identical calculations, the per-cycle cost of `libterachem.py` is equal to or slightly below that of the native executable. Figure 10 decomposes the per-cycle wall time into its component parts. The J, K, and XC builds coincide between the two programs to within run-to-run scatter (as expected, since both call the same GPU kernels). We group the remainder of the cycle as other SCF work, the per-cycle total minus the three builds, which collects the Fock diagonalization, the DIIS solve, the density build, and the per-cycle bookkeeping. On this hardware the remainder is modestly smaller for `libterachem.py` and this difference narrows as the system grows. We note that this margin reflects differences in allocation, data movement, and per-cycle bookkeeping and could be hardware-dependent. The conclusion is not that one approach is better than the other, but rather that any overhead associated with using Python is negligible.

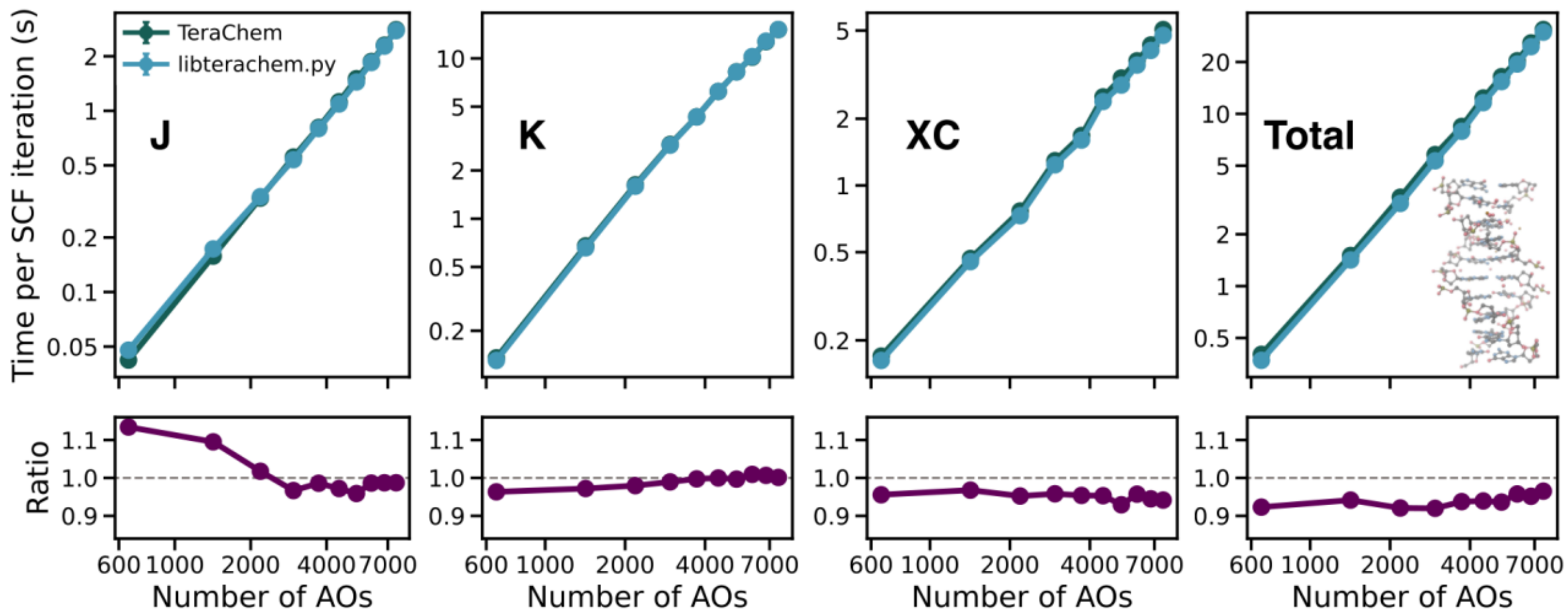

**Figure 10.** Per-SCF-iteration timing of `libterachem.py` and the native TeraChem command line for DNA fragments of one to ten base pairs (655 to 7550 atomic orbitals) on one NVIDIA H100 PCIe 80 GB GPU. Left to right: J, K, XC, and total per-cycle wall time. In each panel the upper axis shows absolute time (log–log) and the lower axis the ratio of timings from `libterachem.py` and TeraChem (which is always close to one).

A composed workflow may also call an external library on every cycle. Figure 11 repeats the comparison with the XC term supplied by an external Python library (PyLibXC) instead of the built-in functional. On the `libterachem.py` side each cycle drives this term from Python. GridBox collocates the density on the quadrature grid, PyLibXC evaluates the functional on the CPU, and GridBox contracts the resulting potential into the Fock matrix. The native command line runs the same LibXC functional through its own in-process adapter. The externalized path is within 6% of the native one and converges toward it with system size. Even though the host-memory array passing of Section II moves the density and potential between the boxes and PyLibXC on every cycle, the overhead associated with the externalized XC path is negligible.

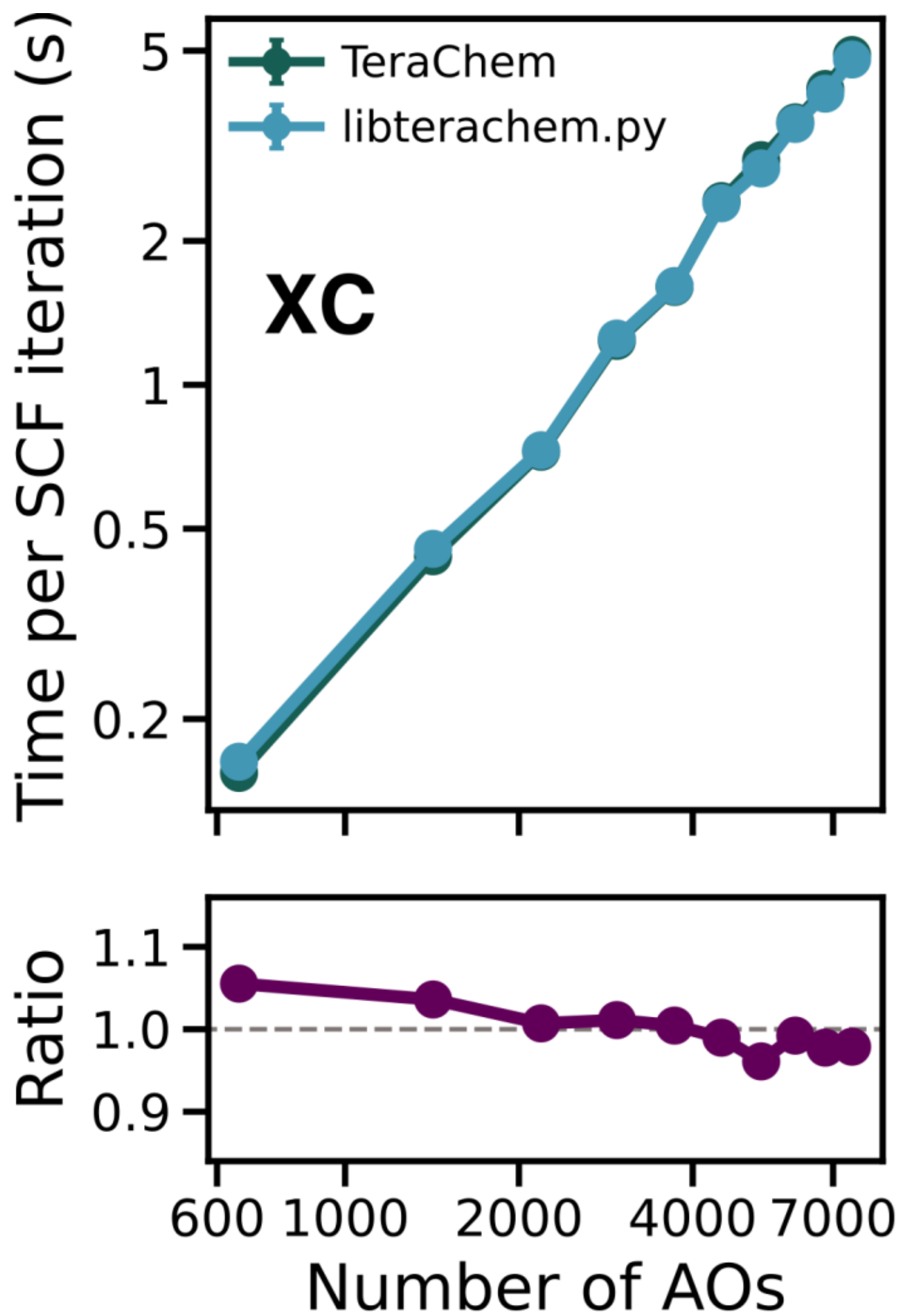

**Figure 11.** Per-cycle XC timing with the functional supplied by PyLibXC (`hyb_gga_xc_wb97x`) rather than the built-in DFTBox functional, for the DNA fragments of Figure 10 on one NVIDIA H100 PCIe 80 GB GPU. Upper axis: absolute XC time (log–log). Lower axis: the `libterachem.py`/TeraChem timing ratio, which remains close to unity.

Because production calculations often spread a single system across several GPUs, Figure 12 runs the ten-base-pair fragment (7550 atomic orbitals) on one to eight GPUs of two contrasting types. The NVIDIA A100-SXM4-40GB is a data-center GPU with fast double-precision arithmetic. The NVIDIA GeForce GTX 1080 Ti is an older consumer GPU with much weaker double-precision throughput. On the A100, libterachem.py tracks the executable at every GPU count, with a per-cycle time of 90% to 93% of the command line. Both programs plateau near a twofold speedup, peaking at five to six GPUs. On the 1080 Ti, a single-GPU cycle takes roughly four times longer, which leaves more work to be distributed among the GPUs. The calculations therefore scale further, reaching speedups of 3.8 (TeraChem) and 3.1 (libterachem.py) on eight GPUs. libterachem.py is faster on up to three GPUs and up to 12% slower beyond. We attribute this crossover to the other SCF work, whose hardware dependence we noted for the single-H100 benchmarks. Nevertheless, libterachem.py closely follows the scaling of the native executable on both GPU types. The contrast also exposes a growing challenge for multi-GPU calculations. As individual GPUs become faster, a fixed-size problem leaves less work for each device, and strong scaling becomes harder to achieve for both programs.

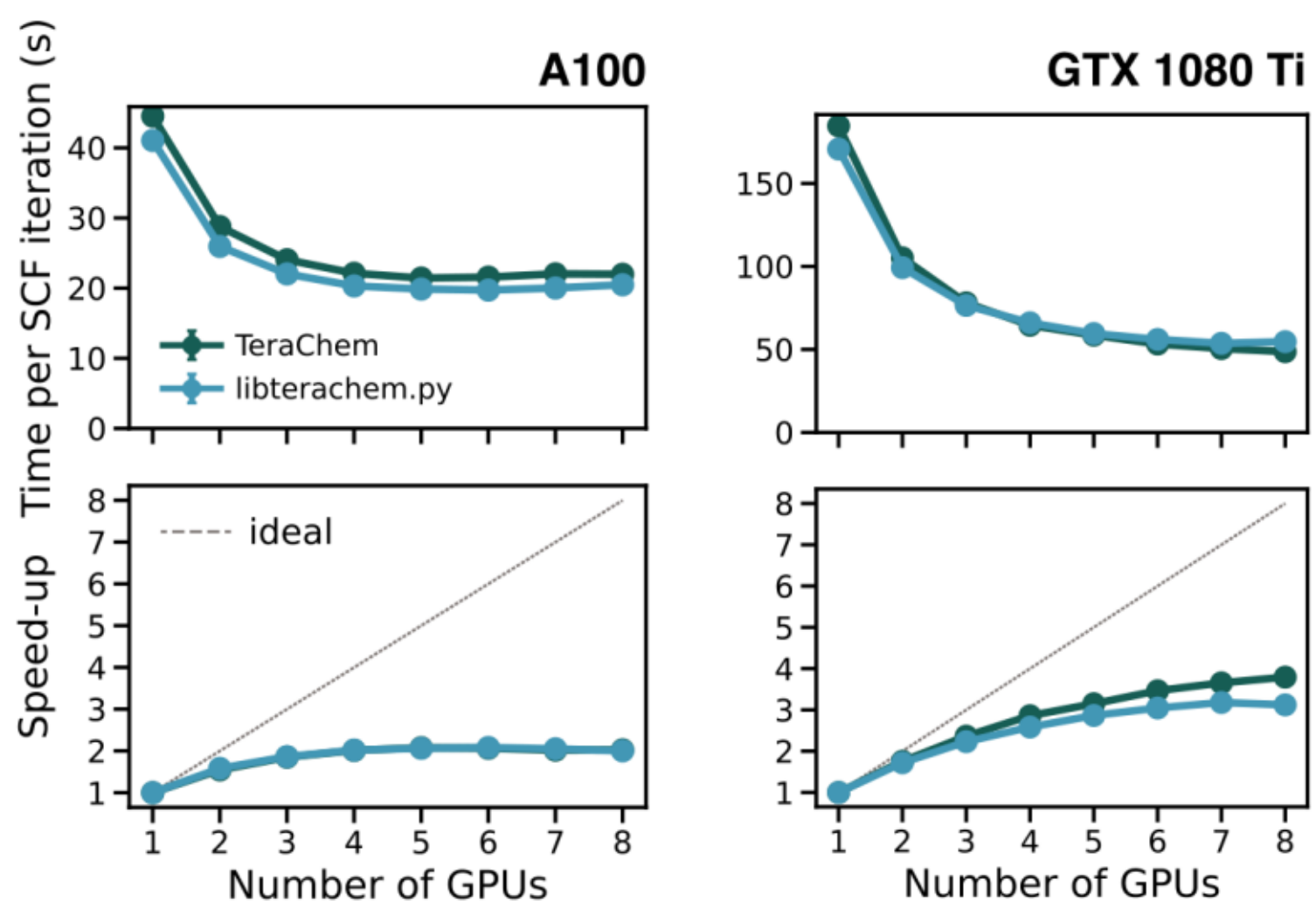

**Figure 12.** Multi-GPU scaling of the SCF cycle for the ten-base-pair fragment (7550 atomic orbitals) at fixed size on one to eight GPUs with the built-in $\omega$B97X functional. Left pair: NVIDIA A100-SXM4-40GB. Right pair: NVIDIA GeForce GTX 1080 Ti (one run per GPU count). Within each pair, the left panel shows the total per-cycle wall time for each program and the right panel the speedup relative to one GPU of the same type ($T_1/T_p$), with ideal linear scaling shown for reference.

***B. QM/MM dynamics and excited-state calculations of PYP***

We demonstrate the combined workflow through ground-state QM/MM dynamics and subsequent excited-state calculations of photoactive yellow protein (PYP), whose chromophore is covalently attached to Cys69.[69-70] Figure 13(a) shows the core setup. The QM/MM layer constructs the QM subsystem and its coupling to OpenMM,[41] which enters the SCF calculation through the same `external_methods` interface introduced in Section III-D. The combined system is exposed to an unmodified ASE integrator,[42] which propagates all 10,935 physical atoms. OpenMM's CUDA platform evaluates the MM contribution alongside TeraChem's GPU-accelerated electronic-structure calculations.

We adapted the published Amber–TCPB benchmark topology[25] and definition of the four nested QM regions containing 21, 46, 97, and 145 physical atoms (22, 49, 104 and 159 including link-atom caps), as shown in Figure 13(b).[71] The simulations used electrostatic embedding without a distance cutoff and a 30 Å spherical boundary restraint. The truncated droplet was first equilibrated for 49 ns of classical molecular dynamics, and its final configuration served as the common starting point for all four regions. B3LYP[72-74]/def2-SVP[75] calculations used 244, 509, 994, and 1479 basis functions. The formal QM-subsystem charge is −1 for QM1/QM2 and zero for QM3/QM4, where Arg52 enters the QM region. Each trajectory started from identical coordinates and initial velocities and underwent 2 ps of Langevin thermalization (NVT) at 300 K, followed by 5 ps of microcanonical (NVE) propagation with a 0.5 fs timestep. The QM1 and QM2 trajectories were continued for a further 5 ps to confirm stationarity. All comparisons use the first 5 ps window common to the four regions.

Figure 13(c) compares end-to-end throughput with TeraChem's built-in `run md` driver on the same hardware and TeraChem build, with matched electronic-structure settings and timestep. At matched GPU counts, the Python workflow achieves 1.05–1.18× the throughput of the built-in driver. Because the MM force is evaluated on the GPU with OpenMM in the Python workflow but on the CPU in the built-in driver, these timings include GPU/CPU differences in the MM evaluation and are not restricted to interface overhead. This is, however, a clear advantage of `libterachem.py`. Because it is awkward (although certainly possible) for TeraChem and OpenMM to use the same GPU, the MM force field is typically evaluated on the CPU when running QM/MM with native TeraChem. With `libterachem.py`, it is straightforward to use the same GPU for both TeraChem and OpenMM.

To illustrate an observable accessible from the trajectories, Figure 13(d) shows the phenolate C4′–O4′ bond length, defined on identical chromophore atoms across all regions. Each region contributes five consecutive 1 ps block means over a matched 5 ps window. The QM1 block means are disjoint from those of QM2–QM4: enlarging the region from QM1 to QM2, which promotes the Tyr42 and Glu46 hydrogen-bond donors into the QM subsystem, lengthens the bond by 0.024 Å, reproduced both in the preceding thermalization and on a doubled window. The further additions at QM3 and QM4 leave the donor geometry unchanged and shift the bond by −0.0005 and +0.004 Å, comparable to or smaller than the within-region block-to-block variation. These are descriptive comparisons from one trajectory per region, not equilibrium uncertainty estimates: the 5 ps windows sample only a limited range of protein and solvent configurations.

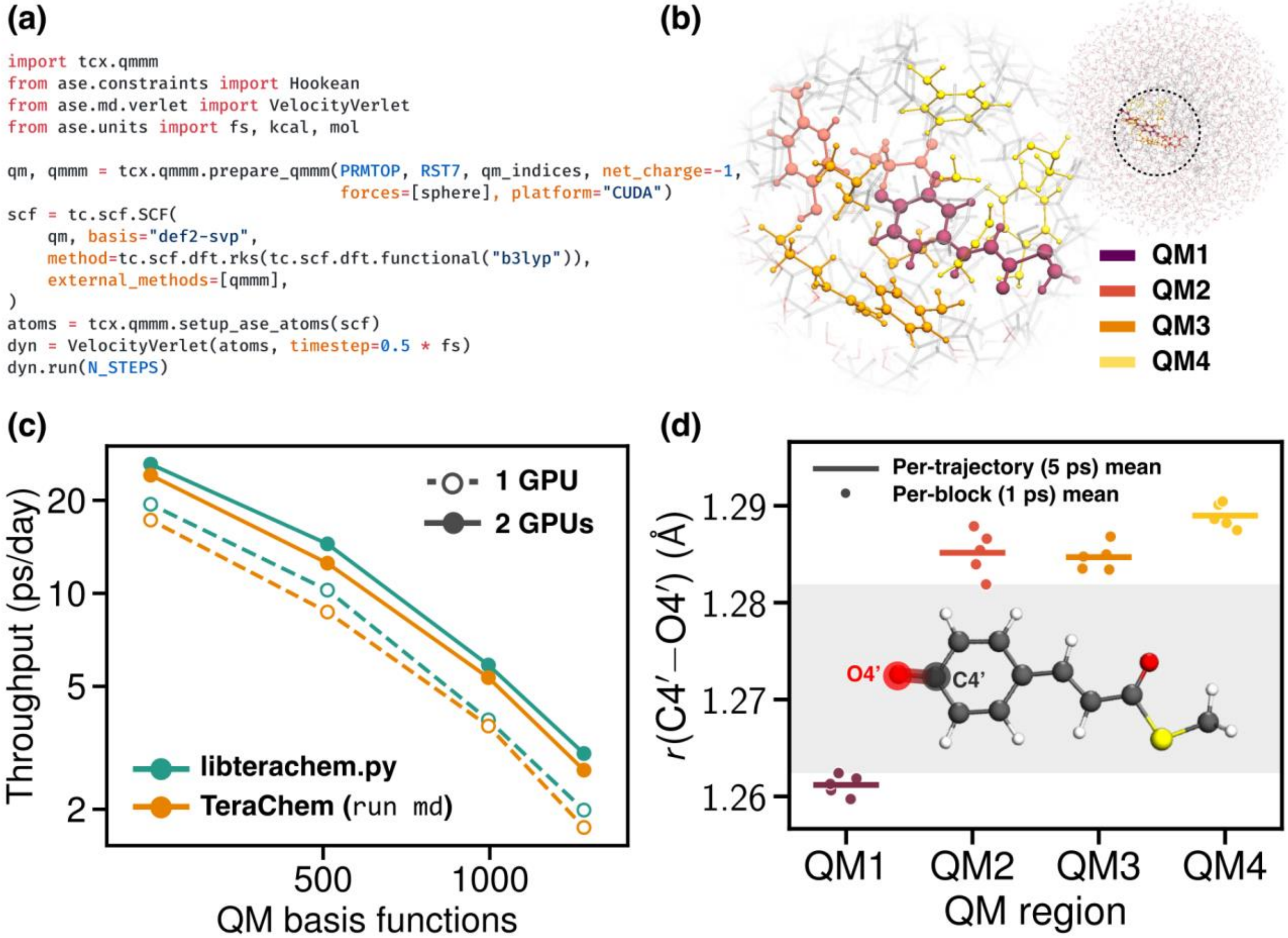

**Figure 13.** QM/MM dynamics and throughput of PYP through `libterachem.py`. (a) Core simulation setup: velocity initialization, thermalization, and trajectory output are omitted. (b) Active-site enlargement and system overview. Colors identify QM1 (the covalently bound *p*-coumaryl chromophore with the Cys69 S–C) and successive additions to the QM region: Glu46/Tyr42 (QM2), Thr50/Arg52/Val66/Tyr98 (QM3), and Phe62/Ala67/Pro68/Thr70/Phe96 (QM4). The MM environment is shown in pale gray. (c) Throughput on NVIDIA A100-SXM4-40GB GPUs. The MM contribution uses OpenMM's CUDA platform in the Python workflow and CPU execution in TeraChem's built-in driver. (d) Phenolate C4′–O4′ bond length for each QM region. Points are means over non-overlapping 1 ps blocks of a matched 5 ps NVE window, five per region, and horizontal bars are per-trajectory means. The shaded band marks a range containing no block mean of any region.

The same QM/MM coupling supports multireference single-point calculations on saved geometries (top of Figure 14). Starting from RHF/def2-SVP, we optimized three equally weighted singlet states using state-averaged CASSCF (SA-CASSCF) with CAS(6,6) and CAS(12,11). The `CASSCF` class is assembled from the box-level operations, taking its configuration-interaction solutions and reduced density matrices from CIBox and its active-space integrals from atomic-orbital J builds in IntBox, following the atomic-orbital-based formulation implemented in TeraChem.[5] Initial orbitals are supplied via a user-provided Python function: it constructs chromophore-centered $\pi$-like combinations within the occupied valence and virtual RHF spaces separately, then selects frontier orbitals by diagonalizing the Fock operator within these subspaces, the same general strategy as the automated $\pi$-orbital space (PiOS) method.[76] The orbitals are optimized by CASSCF, and their chromophore $\pi$ character is checked after convergence. This interface can also accommodate orbital preparation by external Python tools, given compatible basis and orbital conventions.

We evaluated all four QM regions on 15 geometries selected from the QM4 trajectory, holding the nuclear coordinates fixed across regions. Figure 14 (left) shows a stronger QM-region dependence of the mean $S_0 \rightarrow S_1$ excitation energy for CAS(12,11) than for CAS(6,6). Enlarging the region from QM1 to QM4 shifts the CAS(12,11) mean by +0.50 eV, while the CAS(6,6) mean is unchanged within 0.01 eV, illustrating how active-space choice affects the calculated response to QM-region enlargement. We do not attempt to assign or track electronic-state character across geometries or QM regions. Roots are identified solely by their energy ordering.

Figure 14 (right) reports the wall time of one CASSCF macro iteration against QM basis-function count for the two active spaces on an A100 GPU, with the cost of one RHF cycle shown for reference. Reporting per-iteration cost separates the expense of an iteration from the number of iterations a particular geometry happens to need. Both active spaces scale sub-quadratically over the six-fold range in basis-function count. The timings cover the in-process RHF and CASSCF work under a two-step orbital optimizer[77-78] with the same convergence thresholds used above, and exclude GPU start-up and QM/MM construction.

```python
# RHF/QM/MM setup and pi_guess definition omitted.
cas = CASSCF(active_space=ActiveSpace(num_active_orbitals=11, num_closed_orbitals=n_closed, active_orbitals=active_orbs),
             guess_fn=pi_guess, method=DirectCI(root_spaces={1: 3}), converger=TwoStep(conv_thre=1e-5), max_macro_iter=100)
with scf.session() as s:
    s.energy()
    e = cas.run(s).ci_result.state_energies
```

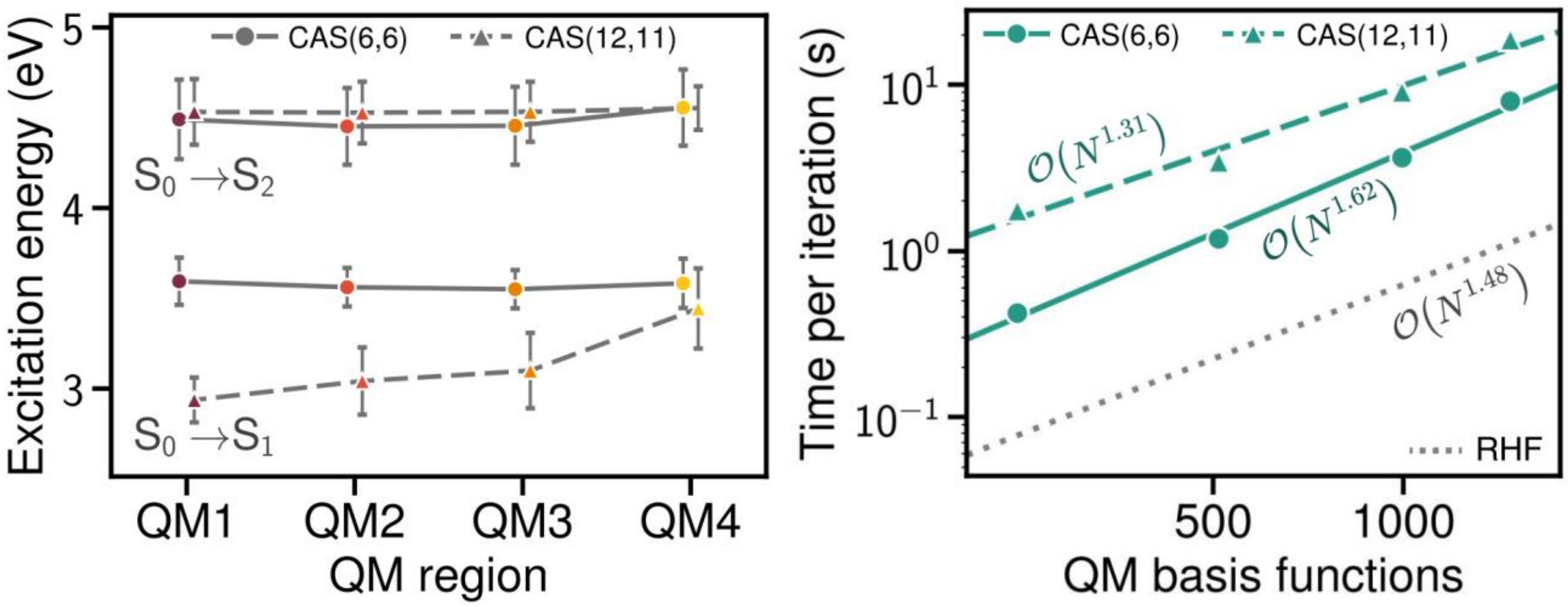

**Figure 14.** SA-CASSCF/def2-SVP calculations of PYP with three equally weighted singlet states. Top: representative CASSCF setup. Left: mean vertical excitation energies over 15 common QM4-trajectory geometries with bars showing ±1 standard deviation across geometries for each QM region and active space. Right: mean wall time per CASSCF macro iteration against QM basis-function count, with the dotted line giving the cost of one RHF cycle. Lines are power-law fits and the annotations give the fitted exponents.

## V. Conclusions

We have presented `libterachem.py`, a Python package that exposes TeraChem's production GPU operations and their intermediate quantities. The `tc` framework assembles these operations into complete methods, and the `tcx` layer composes them with external libraries. Electronic-structure methods can thus be written and modified in Python while the GPU implementations beneath them remain unchanged.

We demonstrated this at both levels of the architecture. At the operation level, we examined grid pruning and fitting-matrix conditioning in least-squares THC, evaluated XC contributions through PyLibXC and the machine-learned Skala functional, and assembled SA-CASSCF from CIBox and IntBox operations. At the method level, complete methods were combined with dispersion corrections, periodic boundary conditions, geometry optimization, and QM/MM molecular dynamics. Driving the SCF iterations from Python comes at a cost comparable to TeraChem's native driver and preserves its multi-GPU scaling. The PYP QM/MM workflow likewise reaches throughput comparable to TeraChem's built-in driver.

In future work, we will extend `libterachem.py` to excited-state electronic structure, spanning linear-response and multireference methods. A central target is excited-state dynamics such as ab initio multiple spawning (AIMS), which requires energies, nuclear gradients, and nonadiabatic couplings for several electronic states at every time step.[79-81] The SA-CASSCF calculations demonstrated here provide a starting point, and a dynamics driver can connect through `tcx` as ASE does for ground-state dynamics.

Beyond the examples presented here, a method built on `libterachem.py` requests its operations through fixed interfaces. Improvements to the underlying kernels and ports to new GPU hardware therefore are automatically usable by every such method, as long as the interfaces remain stable. The same interfaces accept implementations from outside TeraChem, so a method need not draw all of its operations from a single program. We expect GPU electronic-structure operations to become a shared resource across programs, from which a developer selects the best-suited implementation for each step. Our aim is a setting in which new electronic-structure methods are prototyped with the immediacy of Python, run at the speed of optimized GPU code, and easily composed with different programs and libraries in the growing ecosystem of electronic structure software.

**Supplementary Material**

See the supplementary material for the benchmark geometries (the one- to ten-base-pair DNA fragments, the photoactive yellow protein droplet topology, its four nested QM-region definitions, and the shared starting configuration), the raw per-cycle timing data (SCF overhead, the PyLibXC-supplied XC path, multi-GPU scaling, and QM/MM throughput) together with the CASSCF wall times, and the input specifications for all reported calculations. The figure code excerpts are archived in the associated Zenodo repository https://doi.org/10.5281/zenodo.23071219.

**Acknowledgments**

We thank the many past and present TeraChem contributors for their work on the underlying TeraChem code. This work was supported by the AMOS program of the U.S. Department of Energy, Office of Science, Basic Energy Sciences, Chemical Sciences, Geosciences, and Biosciences Division. AEHB acknowledges financial support from the Novo Nordisk Foundation under grant reference number NNF24OC0089345. We thank Martin Stöhr for discussions on incorporating the many-body dispersion example.

## Author Declarations

### *Conflict of Interest*

T.J.M. is a co-founder of PetaChem, LLC.

## Data Availability Statement

The data that support the findings of this study are available within the article, its supplementary material, and the associated Zenodo repository https://doi.org/10.5281/zenodo.23071219, which also archives the figure code excerpts and is released under CC-BY-4.0.

The `libterachem.py` framework (`terachem_box`, `terachem_util`, `tc`, and `tcx`) is available at https://github.com/mtzgroup/libterachem-py and https://github.com/mtzgroup/tcx. The results in this work were produced with framework version 2026.9.0.